# Nanoscale Mapping of Nanosecond Time-scale Electro-Mechanical Phenomena in Graphene NEMS

Nicholas D. Kay†, Peter D. Tovee†, Benjamin J. Robinson†, Konstantin S. Novoselov‡, and Oleg V. Kolosov*†

† Department of Physics, Lancaster University, Lancaster LA1 4YB, UK

‡Department of Physics and Astronomy, Manchester University, Manchester M13 9PL, UK

*Email: o.kolosov@lancaster.ac.uk

**Atomically thin layers of two-dimensional (2D) materials such as graphene, $MoS_2$ and h-BN have immense potential as sensors and electronic devices thanks to their highly desirable electronic, mechanical, optical and heat transport properties[1,2]. In particular their extreme stiffness, tensile strength and low density allows for high frequency electronic devices, resonators and ultra-sensitive detectors[3,4] providing realistic avenues for down-scaling electronic devices and nanoelectromechanical systems (NEMS). Whilst nanoscale morphology and electronic properties of 2D materials can be studied using existing electron or scanning probe microscopy approaches, time-dependant phenomena on the ns and shorter time-scales cannot be readily explored. Here we use the heterodyne principle[5] to** **reach into this ns time-scale and create a local nanoscale probe for electrostatically induced actuation of a graphene resonator, with amplitude sensitivity down to pm range and time sensitivity in the ns range[6]. We experimentally observed response times of 20-120 ns for resonators with beam lengths of 180 nm to 2.5 μm in line with the theoretical predictions for such NEMS devices.**

To understand the fundamental dynamic properties of graphene NEMS a variety of techniques have been adopted such as detecting the motion directly with an atomic force microscope (AFM) probe[7-9], optical probes[10,11] or capacitive/pull-in sensing[12,13] all with their merits and drawbacks. Clearly, the lateral resolution of the far-field optical methods is limited by the wavelength of light to the μm length scale[14] whereas capacitive pull-in only measures the response of overall NEMS structure[15]. The use of AFM provides the nm resolution[7,16] however as NEMS become smaller and, as a result, faster (with frequencies reaching into the GHz range[17]) it becomes impossible for traditional scanning probe techniques to obtain the required information. To overcome this we here implement a new approach that we call electrostatic heterodyne force microscopy, or E-HFM.

E-HFM combines the time-domain sensitivity of heterodyne force microscopy (HFM)[18-20] and the capability of contact electrostatic force microscopy (EFM)[21,22] for probing electromechanical phenomena on the nanometre length scale. In E-HFM an ultrasonic excitation of the tip modulates the relative distance between the tip and the probed NEMS at the frequency $f_t$, thus switching on and off the tip-surface force interaction at frequencies that can range from a few MHz to above 100 MHz[23]. This acts as the equivalent of a "local oscillator" in a heterodyne mixer. By using the heterodyne detection principle[24] we are able to down-convert much smaller (pm range) amplitudes of electrostatically induced vibration from the NEMS at the frequency $f_s$ to much lower and therefore more easily detectable frequency $\Delta f = f_s - f_t$ preserving both the amplitude of NEMS vibration as well as retaining the phase information. As a result, E-HFM is able to measure nanoscale electro-mechanical properties with a sub-ns temporal sensitivity.

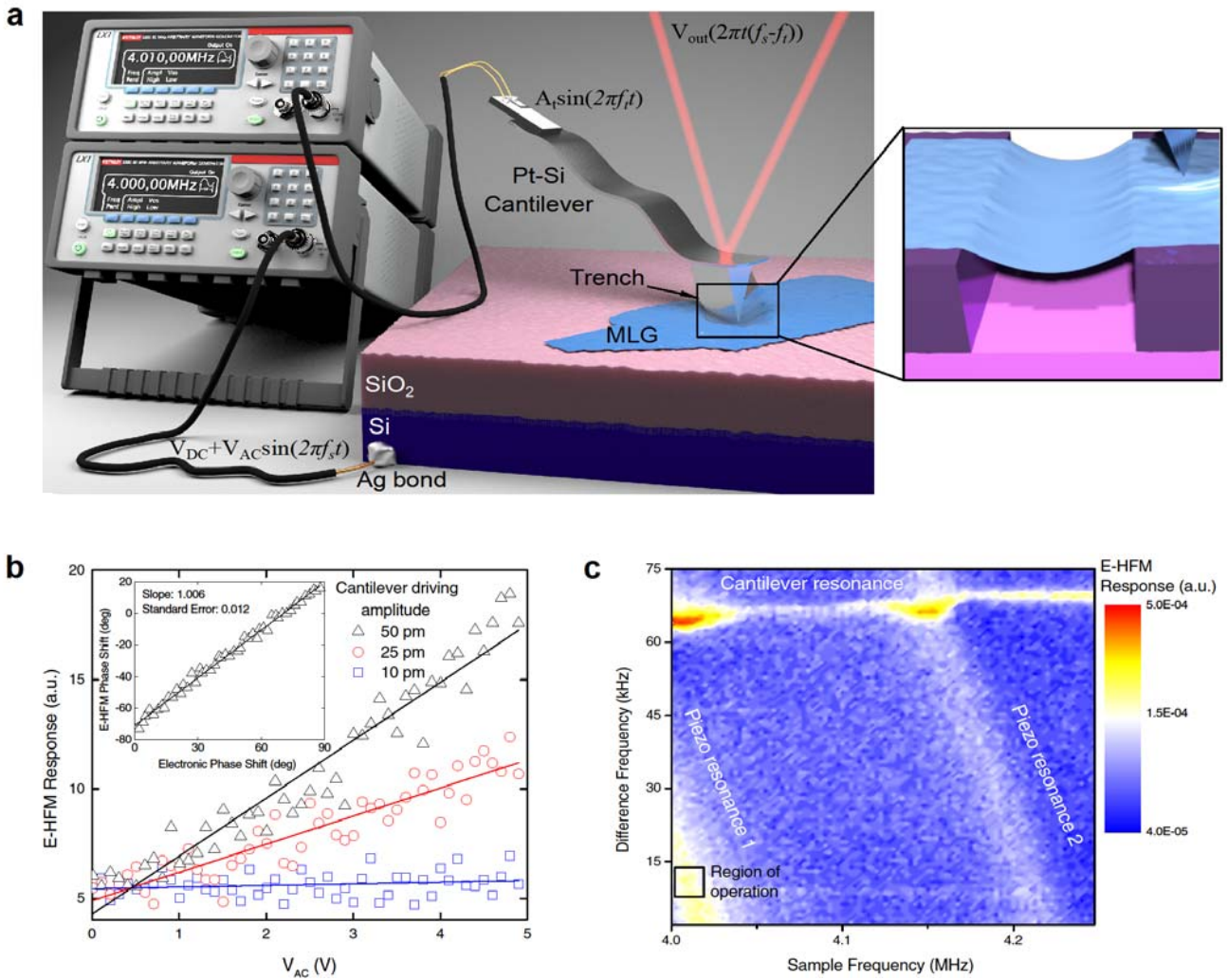

**Figure 1 | Experimental setup and working principle of E-HFM. a** Identical function generators are used to drive the tip and sample at adjacent frequencies; a piezo transducer attached at the base of the cantilever drives the tip mechanically with frequency $f_t$. Simultaneously both the sample and cantilever are also subjected to the electrostatic forces via the bias applied to the back gate with AC frequency $f_s$. A conductive Pt-Si cantilever is used which also grounds the graphene. *Zoomed area* in **a** illustrates a typical suspended beam structure of the 2D-NEMS we studied. **b** E-HFM amplitude response at the difference frequency $\Delta f = f_s - f_t$ as a function of $V_{AC}$ with $V_{DC}$ = 5 V for different peak amplitudes of the cantilever piezo; the linear response to the driving amplitude is clearly seen. The frequencies used are $f_t$ = 4 MHz and $f_s$ = 4.01 MHz. *Inset* in **b** - the E-HFM phase response as a function of an electronic phase shift of one of the function

generators showing a one-to-one phase relationship. **c** The E-HFM amplitude as a function of driving frequency $f_s$ and frequency difference between the electrostatic actuation and a "local oscillator" $\Delta f$. The sample frequency $f_s$ is lower than the tip frequency by $\Delta f$. The voltage applied to the sample was 10 V$_{p-p}$ AC + 5 V DC whilst the cantilever was driven at the base with an amplitude of 50 pm. The region of operation for this study is situated far from a cantilever resonance to provide genuine phase measurements. Both graphs **b** and **c** were produced whilst the tip rested on SiO$_2$ substrate with contact force of 11±3 nN whilst scanning was disabled.

A fully comprehensive model of the E-HFM characteristics requires a numerical solution to the Euler beam equation; however, as in our study we use the low-frequency quasi-static limit to eliminate phase shifts due to cantilever resonances, a simpler model of the force acting on the tip at the difference frequency is adequate (see Supplementary Information Note 1):

$$F_{ts}(t) = \chi A_s A_t \cos\left(2\pi f_s t - 2\pi f_t t + \phi_s - \phi_t\right) - \frac{3 A_t c^2 L w \varepsilon_0 \chi}{2k} V_{AC} V_{DC} \cos\left(2\pi f_s t - 2\pi f_t t + \phi_c - \phi_t\right) \quad (1)$$

Here the symbols $A$, $f$ and $\phi$ denote the amplitude of induced vibration, frequency of the vibration and phase, respectively, whilst subscripts denote whether the vibrations are induced by the tip piezo "$t$", through electromechanical actuation of the cantilever "$c$", or sample "$s$". The electric potential of the sample and the cantilever are the same due to direct contact of the conductive tip and the flake, ensuring that the electromechanical actuation of both occur at the same frequency $f_s$. $V_{AC}$ and $V_{DC}$ are the AC and DC voltage differences between the back-gate and the sample/cantilever.

Equation (1) shows that the E-HFM operation can be attributed to a combination of two mixing phenomena, with the "large" ~1 nm ultrasonically excited amplitude vibration from the tip $A_t \cos\left(2\pi f_t t + \phi_t\right)$ acting as a "local oscillator" which is combined with the sample and cantilever electrostatic actuation resulting in the E-HFM signal. Note that in the conventional HFM[24] there is mixing of only the sample and the tip vibrations, resulting in only the first term being present in the equation (1). E-HFM relies on the "local oscillator" mechanical tip vibration being large enough to cross the non-linear region of the tip-sample interaction described elsewhere[6,18] which is the order of 1 nm, to produce a reliable "switching" on and off of the tip-surface contact. Once this threshold amplitude has been achieved, a vibration of the sample and/or cantilever that is much smaller than the amplitude of the tip vibration will result in a linear increase in the signal at the difference frequency $\Delta f$ – reflecting the heterodyne regime of detection as illustrated in the Fig. 1b. It can be deduced from Fig. 1b that small driving amplitudes (9±5 pm) at the base of the cantilever will correspond to a tip amplitude below 0.2-0.5 nm, this being insufficient to move the tip-sample across the non-linear region and therefore will result in no signal at the difference frequency, however for larger amplitudes (>26±5 pm) the heterodyne regime is achieved. Fig. 1b inset provides assurance that inducing an electronic phase shift at $f_t$ with respect to $f_s$ we observe an equal shift in the phase of the mixed-down E-HFM signal demonstrating that the phase of the high frequency signal is directly translated to the down-shifted frequency $\Delta f$.

In previous studies of "standard" HFM it has been shown that it is possible to tune the difference frequency to various modes of the cantilever to enhance the signal [25,26]. To analyse this, a two-dimensional sweep was performed of the E-HFM response as a function of the tip driving frequency $f_t$ and the difference frequency $\Delta f$. Fig. 1c illustrates two main frequency-dependent mechanisms affecting the E-HFM response - the resonant response of the cantilever piezo driving the tip and the contact resonance of the cantilever itself. By combining the cantilever contact resonance with one of the piezo resonances one can achieve a large increase in response. With the cantilever contact resonance measured as Q~30 even relatively small sample amplitudes of 10 pm would result in a detected amplitude in the region of 0.3 nm. While that may seem to be useful, operating E-HFM close to the contact resonance inevitably affects the signal phase, with local sample mechanical properties that may additionally shift the resonance inducing false amplitude and phase contrast. That prompted us to work in low, non-resonant $\Delta f$ frequencies (Fig. 1d) throughout this study.

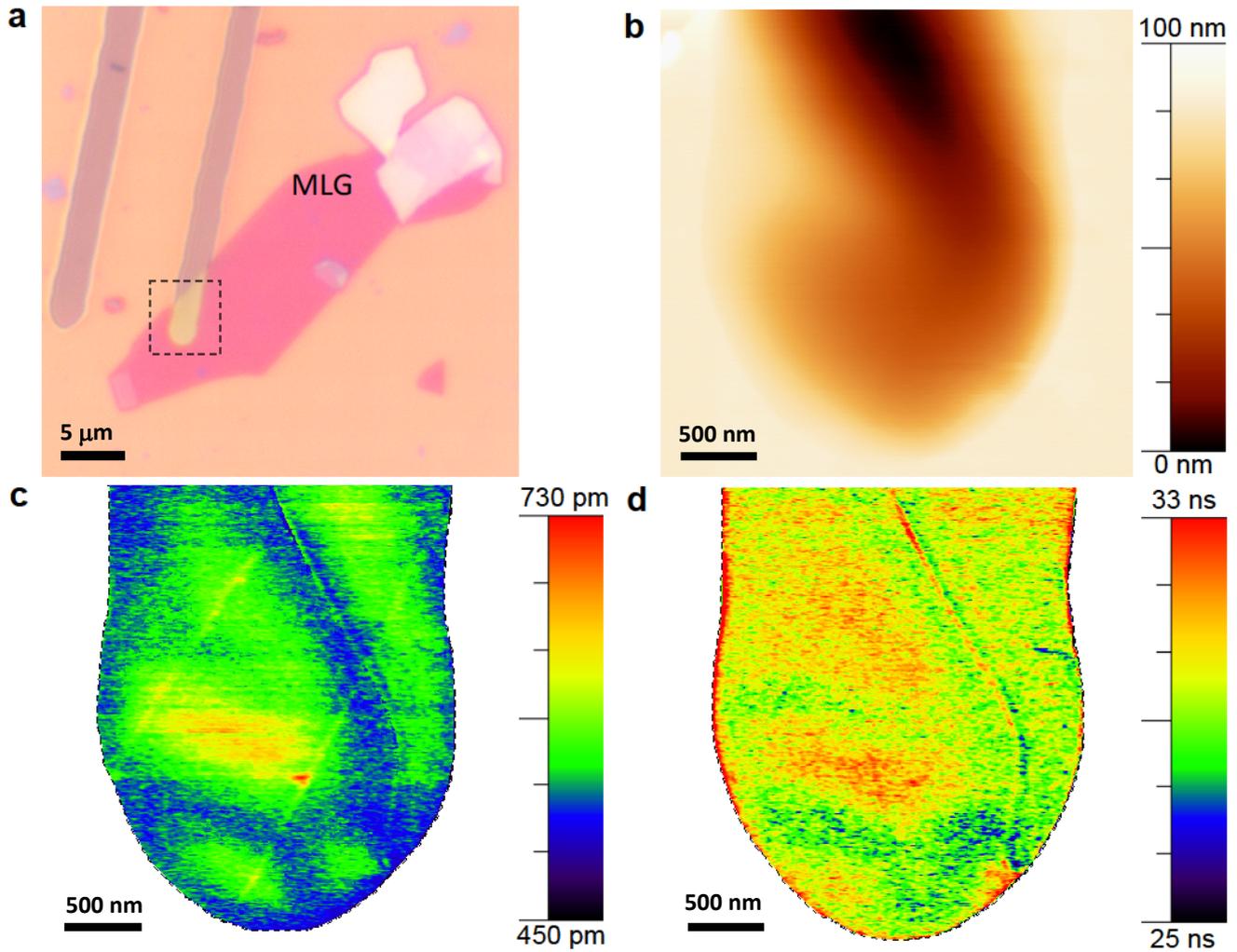

**Figure 2 | E-HFM mapping of the amplitude and time-domain response of a multi-layer graphene (MLG) flake. a** Optical microscopy image of the 7 nm thick MLG sample studied with the area of interest (dashed box), scale bar 5 µm. **b** AFM topography of the suspended region taken simultaneously to images **c** and **d** with a set force of 0 nN. **c** Amplitude and **d** response time maps of the E-HFM electromechanical response of the flake, scale bar in b-d) is 500 nm, only the suspended region is considered in **c** and **d** and an artificial white background is applied.

E-HFM was then used to study various resonator type NEMS consisting of graphene layers suspended over a 2.5 µm wide trench. Fig. 2c,d shows the spatial variation of the electromechanical properties of the graphene NEMS with Fig. 2c showing the amplitude variation, this signal being a convolution of mechanical stiffness variations and electromechanical actuation of the sample. The pattern observed is most likely linked to the intrinsic stress present in the MLG sample which is correlated with the topographical "kink" observed in Fig. 2b. The E-HFM phase image (Fig. 2d) directly reflects the phase delay of the response at the tip excitation frequency of 4 MHz[20,24] that, if expressed in terms of a fraction of the 250 ns period, results in approximately 8 ns total span in this image. It can be noted that the locally increased E-HFM amplitude in Fig. 2c corresponds to the local changes in the phase shift in Fig. 2d.

In order to understand the time-dependent dynamics of the electromechanical response of graphene NEMS in E-HFM it is necessary to decouple the electromechanical response from mechanical effects. This was done by comparing E-HFM with HFM[24], a purely mechanical heterodyning technique. A comparison of E-HFM and HFM response across the same line is seen in Fig. 3.

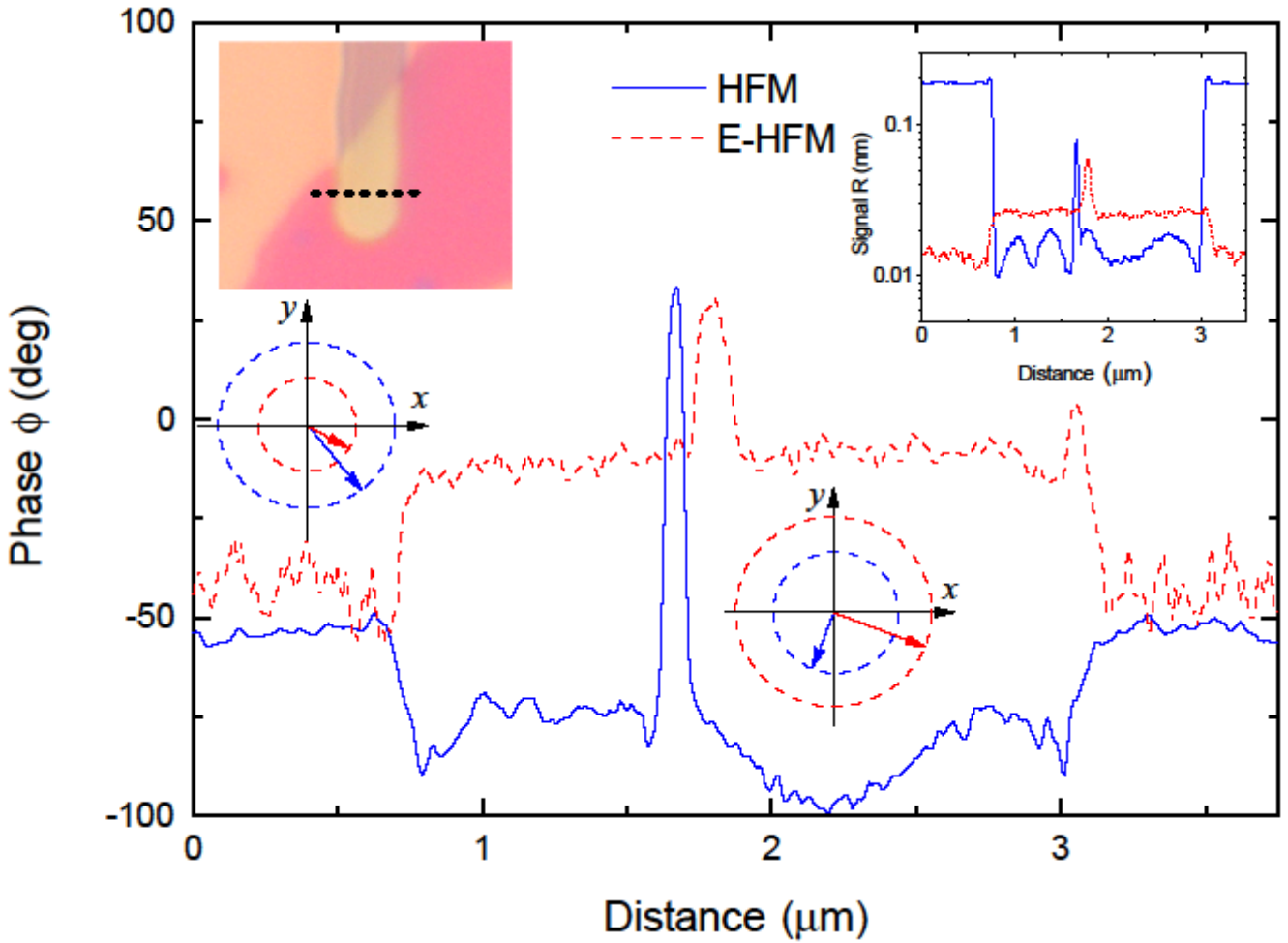

**Figure 3 | Decoupling of electromechanical and purely mechanical actuation by comparison between E-HFM and HFM responses.** Image shows an average trace of the E-HFM and HFM phase signal across a MLG flake suspended over a 2.5 μm trench (left inset). The lateral dimension of the central feature in the profile is on the order of 50 nm. The right inset shows the amplitude of E-HFM and HFM signals and both amplitude and phase of the signals were used to interpret the phase delay produced by the suspended graphene layer.

To decouple the electrostatic from the mechanical effects we compare the E-HFM signal amplitude and phase with that of conventional HFM and consider the phasor diagrams of the cantilever and the NEMS membrane and their contributions to the total signal. We consider the amplitude, and phase of the HFM and E-HFM signals as real components of a phasor $\text{Re}\{R \cdot e^{i\phi} \cdot e^{i\omega t}\}$ considering cases when the tip is in contact with MLG on $SiO_2$ and suspended MLG. We use subscripts "$_E$" and "$_H$" to denote E-HFM or HFM, and "$_S$" and "$_G$" for the tip located on MLG on substrate or free suspended graphene, respectively. The vector calculations reflect that the cantilever and MLG contributions are summed as phasors to produce the amplitude and phase E-HFM response and show that the phase $\phi_{em}$ is associated with the electro-mechanical phenomena (see derivation in Supplementary Information Note 1)

$$\phi_{em} = \arctan\left( \frac{R_{ES}\sin\phi_{ES} - R_{HS}\sin\phi_{HS} + R_{HG}\sin\phi_{HG} - R_{EG}\sin\phi_{EG}}{R_{ES}\cos\phi_{ES} - R_{HS}\cos\phi_{HS} + R_{HG}\cos\phi_{HG} - R_{EG}\cos\phi_{EG}} \right) \quad (2)$$

The response time is then extracted as $\tau_{em} = 2\pi\phi_{em}/f_s$. This method of the addition of phasors assumes that the effect of the sample stiffness on the HFM/E-HFM signal is in the first approximation similar regardless of whether the sample is driven by a piezo (HFM) or if the tip is driven by capacitive forces (E-HFM).

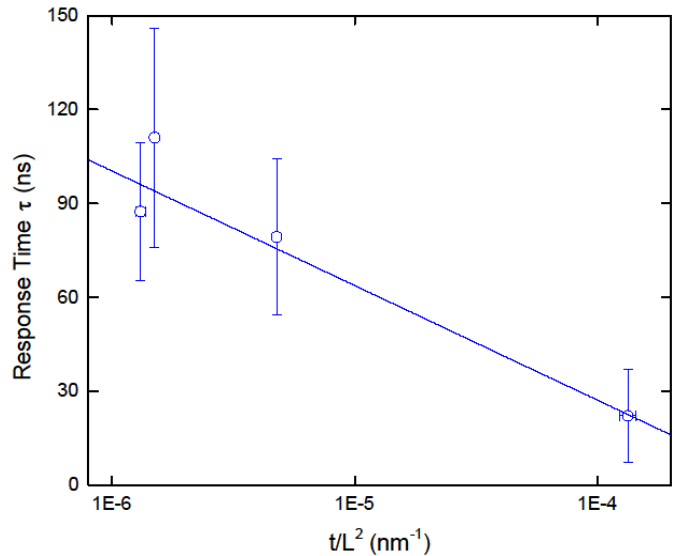

**Figure 4 | The response time dependence on 2DM beam dimensions.** The characteristic response time associated with several 2D-NEMS as a function of $t/L^2$. All response times were acquired at an operating voltage of 10V AC and a DC offset of +5 V.

Figure 4 shows the linear decrease in response times taken at the centre of the NEMS structure. The times observed agree, within experimental error, with the theoretically predicted numbers from the out-of-resonance response of oscillating resonator[27]. From this we deduce that the physical mechanism of the response time is due largely to the inertial response of the layers to the electrostatic field. The values obtained in this study appear consistent, within a few hundred ns, with data obtained for similar systems in other studies[28]. The measured differences are thought to be attributed to a variety of factors that are not accounted for such as intrinsic beam tension due to differing manufacturing procedures, buckling of the 2DM membrane[15] as well as damping effects on the beam[29].

The demonstrated exploration of time domain electromechanical phenomena in graphene NEMS via E-HFM can be expanded to heterostructures of 2DM's that possess specific electronic, piezoelectric and optoelectronic properties, and their combination with traditional NEMS structures like Si and silicon nitride membranes[30]. The inherently low noise and linear detection mechanism of heterodyne approach will allow detection of ultra-low amplitude vibrations down to few tens of fm scale, whereas the driving frequency of mechanical heterodyning can be easily expanded to multi-GHz frequencies by using, eg SAW approach, bringing the time scale of sensing phenomena to the ps range.

**Methods**

**Sample substrate preparation.** The trench etch mask was produced with an optical mask and a single layer of S1813 resist spin coated to a thickness of 1.6 µm. The pattern was etched into Si/$SiO_2$ wafers (300 nm oxide) using a plasma mixture of $CHF_3$ (35 sccm) and $O_2$ (10 sccm) at a pressure of 10 mTorr and an etch rate of 0.2 nm/s producing trenches of 150 nm in depth. This etch depth ensured that the deposited flakes were fully suspended but also left enough dielectric so that the tip did not short out with the back gat (see also Methods in Supplementary Information)  **E-HFM Signal Detection.** The E-HFM signal is detected using a Stanford Research Systems SR830 DSP Lock-In Amplifier where the deflection signal of the AFM (Bruker MultiMode® 4) is used as the input. The reference signal for the lock-In is provided by using an RF mixer where reference signals from the two Keithley 3390 function generators are provided as the LO and RF inputs, an additional 100 kHz RC filter is used to filter out higher mixing contributions from the two signals. The lock-In amplifier used had a typical phase resolution of 0.1 ° corresponding to a temporal resolution of approximately 70 ps. **E-HFM Cantilever Preparation and Sample Bonding.** The sample is bonded with phenyl salicylate (melting point: 41.5 °C) to a glass slide mounted onto a piezo-electric transducer with a resonance frequency of 4 MHz, this serves the purpose of providing a vibrations at the sample for HFM. To provide an electrical bias to the sample native oxide is removed from the side of the Si/SiO$_2$ wafer and bonded to an insulated Cu wire with Ag paint. The cantilever is also bonded to a custom made PZT piezoelectric transducer with Phenyl salicylate which is in turn mounted on a custom built AFM tip holder. To ensure that the tip is grounded, a small insulated Cu wire is attached to the base of the cantilever chip with Ag paint this then connected to the chassis of the function generator providing the AC+DC bias to ensure there is minimal unwanted load in the circuit.

**Acknowledgements**.

Authors are grateful to Volodya Falko for the valuable discussions on the experimental setup and interpretation of the results, Dagou Zeze and Mark Rosamond for the insight into processing of the trench substrates, and EU support on the grants FUNPROB (GA269169), QUANTIHEAT (GA604668) as well as EPSRC grant (EP/K023373/1).

**Author contributions**

O.V.K suggested the general idea of E-HFM approach, O.V.K, K.S.N. and N.D.K. explored the feasibility of the approach in the application to 2D materials, N.D.K., P.D.T, O.V.K and B.J.R carried out experiments, N.D.K, O.V.K and K.S.N. analysed the data and wrote the manuscript.

**Additional information**

Supplementary information is available in the online version of the paper. Reprints and permissions information is available online at www.nature.com/reprints. Correspondence and requests for materials should be addressed to O.V.K.

**Competing financial interests**

There is no declared financial interstes.

# Supplementary Information

# Nanoscale Mapping of Nanosecond Time-scale Electro-Mechanical Phenomena in Graphene NEMS

Nicholas D. Kay[†], Peter D. Tovee[†], Benjamin J. Robinson[†], Konstantin S. Novoselov[‡], and Oleg V. Kolosov*[†]

**Methods.**

**Trench & Hole Etching.** Trenches were etched into a plain Si-SiO$_2$ substrate with a 300 nm thermally grown oxide. A photo-mask was used to pattern the positive photoresist (S1813, thickness: 1.55 μm) with an array of holes of radius 1.5 μm and trenches with widths of 3 and 2 μm. The optical exposure dosage was approximately 70 mJ cm$^{-2}$. After exposure a "hard" bake was performed at 90°C for 90s to increase the resistance to the plasma etching. The pattern was etched into Si/SiO$_2$ wafers (300 nm oxide) using a plasma mixture of CHF$_3$ (35 sccm) and O$_2$ (10 sccm) at a pressure of 10 mTorr and an etch rate of 0.2 nm s$^{-1}$ producing trenches of 150 nm in depth. The SiO$_2$ substrate was then etched with a plasma mixture of CHF$_3$ and O$_2$ for a total time of 20 minutes to etch 200 nm into the SiO$_2$.

**Deposition of Graphene Flakes.** After etching of the substrate was completed it was cleaned with acetone (5 mins, 50°C) and IPA. Following this the substrate was treated with an $O_2$ plasma to remove any remaining hydrocarbons present on the surface. To deposit graphene/FLG onto the substrate small graphite flakes were deposited onto Loadpoint© 6034 medium tack tape. After cleaving the graphite several times it was brought in contact with the substrate in such a way as to reduce the amount of bubbles present and therefore maximising surface contact. The tape was left on the substrate for a total of 15 minutes before peeling it off slowly so that a maximum amount of material was deposited. Once the presence of a desirable flake was found, the sample was then baked in vacuum for 8 hours at 190°C to remove any water trapped beneath the surface. Removal of this water has been proven critical to gaining an accurate measurement of the number of layers as well as removing any p-type doping.

## Note 1. Derivation of the E-HFM Signal Response

To derive a function for the force acting on the cantilever at the difference frequency we first consider the tip-surface interaction which acts as our non-linear mixer. For this we consider the most basic interaction

$$F = kz + \chi z^2 \qquad \text{Eq. S1}$$

Where *z* is the distance between tip and sample, *k* is the linear coefficient and $\chi$ the non-linear coefficient. To obtain the force we must calculate the distance *z(t)* which will be a combination of the motions from the sample, cantilever and piezo actuator.

The amplitude of the piezo attached to the tip as a function of time *z$_t$(t)* is written as

$$z_t(t) = A_t \sin(2\pi f_t t + \phi_t) \qquad \text{Eq. S2}$$

Where *A$_t$* denotes the amplitude of the tip piezo, *f$_t$* the frequency and $\varphi_t$ the phase of the tip vibration. In a similar fashion the motion of the sample due to electrostatic actuation is

$$z_s(t) = A_s \sin(2\pi f_s t + \phi_s) \qquad \text{Eq. S3}$$

Here we are only concerned with the vibration amplitude, phase and frequency so it is not necessary to consider the beam dynamics of the suspended sample.

The third contribution to the tip-surface distance is the electrostatic actuation of the cantilever. We need to know the displacement of the tip resulting from the electrostatic forces acting on the cantilever. To know the displacement of the tip we to know at which point on the tip surface interaction the tip sits. To do this we express the tip motion due to the capacitive forces as $z_E$ and write the tip-surface interaction as

$$F = k(z_0 - z_E - z_t + z_s) + \chi(z_0 - z_E - z_t + z_s)^2 \qquad \text{Eq. S4}$$

Where $z_0$ denotes the static displacement of the cantilever arising from the set-force. Therefore all other displacements will be subtracted from this except for the $z_s$ which is attracted towards the back-gate and away from the tip. By setting equation Eq. S4) equal to the electrostatic forces acting on the cantilever. The electrostatic force acting on the tip $F_T$ is different from that acting on the cantilever $F_{ES}$ in the following way

$$F_T = F_0 + \frac{3}{8} F_{ES} \qquad \text{Eq. S5}$$

$F_0$ is the static deflection or set-force. The factor of 3/8 comes from beam mechanics of distributed loads acting at the end of a cantilever. By equating equations (Eq. S4 & S5) we are able to rearrange in terms of $z_E$ as follows

$$z_E = \frac{k + 2A_t \chi \sin[2\pi f_t t + \phi_t] + 2A_s \chi \sin[2\pi f_s t + \phi_s] - \sqrt{k^2 + 4\chi(\frac{3F_{ES}}{8} + F_0) + 2\chi z_0}}{2\chi} \qquad \text{Eq. S6}$$

By taking a binomial expansion of the root term in $z_E$ and inserting the whole of $z_E$ into equation (Eq. S4) we obtain an equation with a large number of terms. By reducing many of them trigonometrically and select only those at the difference frequency we arrive at

$$F_{\Delta f} = \chi A_s A_t \cos(2\pi \Delta f t + \phi_s - \phi_t) - \frac{3A_t c^2 L w \varepsilon_0 \chi}{2k} V_{AC} V_{DC} \cos(2\pi \Delta f t + \phi_c - \phi_t) \qquad \text{Eq. S7}$$

Where $\Delta f = f_s - f_t$. For equation (Eq. S7) $F_{ES}$ has merely been replaced by the force on each plate of a parallel capacitor.